\documentclass[notitlepage,10pt,aps,prd,twocolumn,amsmath,amssymb,floatfix,superscriptaddress,nofootinbib,preprintnumbers]{revtex4-1}

\usepackage{amsfonts,amssymb,amsmath,graphicx,color,bm,epsfig}
\usepackage[normalem]{ulem}
\usepackage{multirow}
\usepackage{graphicx}
\usepackage{enumerate}
\usepackage[dvipsnames]{xcolor}
\usepackage[utf8]{inputenc}

\definecolor{ultramarine}{rgb}{0.07, 0.04, 0.56}
\definecolor{cadmiumgreen}{rgb}{0.0, 0.42, 0.24}
\definecolor{indigo(dye)}{rgb}{0.0, 0.25, 0.42}
\usepackage[linktocpage=true]{hyperref}
\hypersetup{
colorlinks=true,
citecolor=ultramarine,
linkcolor=cadmiumgreen,
urlcolor=indigo(dye),
}

\usepackage{xcolor}

\begin{document}
 \hfill         \phantom{xxx}  EFI-18-22

\title{Swampland Conjectures and Late-Time Cosmology}

\author{Marco Raveri}
\affiliation{Kavli Institute for Cosmological Physics, Department of Astronomy \& Astrophysics, Enrico Fermi Institute, The University of Chicago, Chicago, IL 60637, USA}
\author{Wayne Hu}
\affiliation{Kavli Institute for Cosmological Physics, Department of Astronomy \& Astrophysics, Enrico Fermi Institute, The University of Chicago, Chicago, IL 60637, USA}
\author{Savdeep Sethi}
\affiliation{Enrico Fermi Institute \& Kadanoff Center for Theoretical Physics, \\ The University of Chicago, Chicago, IL 60637, USA}
\begin{abstract}
We discuss the cosmological implications of the string swampland conjectures for late-time cosmology, and  test them against a wide range of state of the art cosmological observations. The refined de Sitter conjecture constrains either the minimal slope or the curvature of the scalar potential, and depends on two dimensionless constants. 
For constants of size one or larger, tension exists between observations, especially the Hubble constant, and the slope and curvature conjectures at a level of $4.5\sigma$ and $2.3\sigma$, respectively. Smaller values of the constants are permitted by observations, and we determine upper bounds at varying confidence levels. 
We also derive and constrain the relationship between cosmological observables and the scalar field excursion during 
the acceleration epoch, thereby testing the distance conjecture.
\end{abstract}

\maketitle

\section{Introduction} \label{Sec:Introduction}
There is currently a vibrant debate in string theory about whether space-times with positive cosmological constant are compatible with quantum gravity. If metastable de Sitter space-times are part of the swampland, namely the set of backgrounds that are incompatible with quantum gravity~\cite{Vafa:2005ui}, then the implications for dark energy and late-time cosmology are quite striking. 
Specifically the observed dark energy (DE) must be time-dependent. 

What we know today is that four-dimensional or higher de Sitter space-times are ruled out as solutions of the D=10 or D=11 fundamental supergravity theories~\cite{Gibbons:1984kp, Maldacena:2000mw}, and as solutions of type I/heterotic supergravity together with the leading higher derivative couplings~\cite{Green:2011cn, Gautason:2012tb}, and as solutions of heterotic world-sheet conformal field theory~\cite{Kutasov:2015eba}. This is compelling evidence that de Sitter space-time cannot be found in regions of parametric control in string theory. This same conclusion can also be argued from entropy considerations in regimes of weak coupling in string theory~\cite{Ooguri:2018wrx}. 

On the other hand, there most definitely exist landscapes of supersymmetric flux vacua in string theory with Minkowski space-times. These landscapes were originally constructed in F-theory/type IIB string theory~\cite{Dasgupta:1999ss}, but duality leads to similar supersymmetric landscapes in the heterotic and type I strings. It is worth stressing that these landscapes are perturbative constructions that can certainly be destabilized by non-perturbative quantum effects. 

There are an enormous number of such F-theory backgrounds. Each background is constructed from a given elliptic Calabi-Yau $4$-fold together with a choice of compatible flux and branes, subject to a charge tadpole condition~\cite{Becker:1996gj, Sethi:1996es}. Recent estimates of the number of compactification geometries provide lower bounds of $O(10^{755})$~\cite{Halverson:2017ffz}, and of $O(10^{3000})$ from a recent Monte-Carlo based estimate~\cite{Taylor:2017yqr}. On the other hand, a single given geometry has been estimated to support of $O(10^{272,000})$ distinct flux vacua~\cite{Taylor:2015xtz}. 

The above statements are largely without controversy. The issue of turning the enormous complexity of Minkowski flux vacua into a landscape of metastable de Sitter solutions is far more controversial. The most popular approaches are based on type IIB flux backgrounds which break supersymmetry~\cite{Giddings:2001yu}. Quantum corrections to the low-energy effective action are estimated as if such backgrounds are static solutions of string theory. Unfortunately, such backgrounds are not static solutions of string theory~\cite{Sethi:2017phn}. Quantum effects in string theory, particularly non-perturbative effects but even loop corrections, have to be computed around a meaningful solution of string theory. 

Currently no such time-dependent solution is known. If any solution could be constructed from that initial value data, it is likely to be strongly coupled in either the far future or the far past. The structure of quantum corrections to the space-time effective action would require an understanding of that strongly coupled background. This is the basic problem with type IIB landscape proposals. For related comments as well as a different perspective, see~\cite{Danielsson:2018qpa, Kachru:2018aqn}. Under the assumption that this fundamental problem can somehow be evaded, there are many additional issues concerning uplifting type IIB constructions to de Sitter space-time reviewed in~\cite{Danielsson:2018ztv}, with some very recent analysis found, for example, in~\cite{Bena:2018fqc, Gautason:2018gln, Moritz:2018ani}. 

Conspicuously absent in the preceding discussion is any mention of type IIA or M-theory landscapes of vacua. In both these cases, even the basic ingredients for evading the supergravity no-go theorems are poorly understood. Duality certainly suggests that those ingredients should exist, but the analogues of the higher derivative contributions to both the type IIB tadpole conditions and equations of motion are more complicated; for the tadpole, the contributions are determined by both the choice of flux and metric rather than metric alone~\cite{McOrist:2012yc}. There are interesting attempts to stabilize moduli and construct de Sitter landscapes for compactifications of M-theory on $G_2$ manifolds without flux~\cite{Acharya:2007rc, Bobkov:2009za}. However, it seems likely that flux will again be essential for understanding the structure of generic compactifications in this corner of string theory.      

The other approach has been to propose constructions in massive IIA. The older approaches use large volume Calabi-Yau manifolds as starting points~\cite{DeWolfe:2005uu}. These approaches fail to solve the equations of motion of massive IIA~\cite{McOrist:2012yc}. There are attempts to rescue such approaches by using ingredients like smeared orientifolds. However, orientifolds are defined as quotients of weakly coupled string theory, and they are not smeared. For recent discussions of this and related type IIA issues, see for example~\cite{Escobar:2018rna, Junghans:2018gdb, Banlaki:2018ayh, Andriot:2018ept, Blaback:2010sj, Saracco:2012wc, Gautason:2015tig}. Very recently, de Sitter solutions of massive IIA have been proposed without smearing orientifolds~\cite{Cordova:2018dbb}. The status of these de Sitter constructions will depend on whether one can make sense of O8-planes in a theory like massive IIA, which does not have a perturbative world-sheet description. 

Given the murky status of de Sitter constructions in string theory today, one could adopt one of the following viewpoints:
\begin{enumerate}[(a)]
    \item There is sufficient complexity in the space of string vacua and sufficient ingredients that a landscape of de Sitter solutions, although hard to exhibit, is inevitable.
    \item De sitter space-time is part of the swampland, and dark energy must be time-dependent. 
    \item We do not have enough theoretical understanding yet to make a determination. 
\end{enumerate}
This work is concerned with possibility (b), which has been codified in the swampland conjectures~\cite{Obied:2018sgi, Garg:2018reu, Ooguri:2018wrx}, with further discussion found in~\cite{Dasgupta:2018rtp}, and an alternative conjecture found in~\cite{Andriot:2018mav}. The first of these conjectures provides a simple and powerful constraint on the scalar potentials that can emerge from string theory. It is a bold and provocative claim with observational consequences that merits serious investigation. 
The second conjecture is far less provocative with far more theoretical support, and constrains the validity of effective field theory for large scalar field excursions:  

\begin{itemize}
\item {\bf C1}: The refined dS conjecture requires that any scalar field potential from string theory obeys either, 
\begin{align}
&{\bf C1.1}: \hspace{0.3cm} M_P \frac{|V'|}{V} \equiv \lambda \gtrsim O(1) \,,  \nonumber \\
&\hspace{0.3cm} \mbox{{\bf or}}  \\
&{\bf C1.2}: \hspace{0.3cm} -M_P^2 \frac{V''}{V} \equiv c^2 \gtrsim O(1) \,.  \nonumber
\end{align}
\item {\bf C2}: The distance conjecture constrains field excursions to be small in Planck units over cosmic history if one wishes to trust effective field theory, 
\begin{align}
\frac{\Delta \phi}{M_P} \equiv d \lesssim O(1) \,.
\end{align}
\end{itemize}

Whatever constitutes dark energy, it must behave quite closely to a pure cosmological constant with $\lambda=c=0$, and we want to determine whether the swampland conjectures are already in tension with observation. The main alternative to pure vacuum energy is some version of quintessence~\cite{Agrawal:2018own}.  See~\cite{Colgain:2018wgk}\ for a different dark energy model with a turning point in $H(z)$, and~\cite{DAmico:2018mnx}\ for a strongly-coupled monodromy scenario satisfying the swampland conjectures. Quintessence models are relatively easily embedded in supergravity~\cite{Chiang:2018jdg}, but are much harder to construct in string theory; see, for example,~\cite{Cicoli:2018kdo}. Such models are also accompanied by a host of well-known phenomenological problems; for a very recent discussion and references, see~\cite{Hertzberg:2018suv}.  
In the absence of detected deviations from a $\Lambda$CDM cosmology, cosmological observations will place upper limits on the constants $\lambda$ and $c$ involved in these conjectures.  We will test the classes of potentials that place the weakest bounds on these quantities to arrive at the most conservative assessment of these conjectures. 

While observational bounds on $\lambda$ have been recently examined in the context of C1.1~\cite{Agrawal:2018own,Heisenberg:2018yae,Akrami:2018ylq}, we complete the study of the observational viability of the C1 conjecture with an assessment of $c$ as well.  We also carefully address the dependence of the constraints on the data employed, especially the Hubble constant, and treat both background and linear perturbation observables exactly rather than approximately or through proxy statistics. 
Finally we determine quantitative observational bounds on field excursion both in conjunction with C1, and in the context of C2 alone using dark energy reconstruction techniques.

This paper is organized as follows: in  Sec.~\ref{Sec:PotentialsExcursion} we discuss the potentials involved in testing the swampland conjectures and their implications for field excursion;
in Sec.~\ref{Sec:MethodData} we detail the cosmological data sets that we use to obtain the constraints presented in Sec.~\ref{Sec:Results}. We summarize our findings in Sec.~\ref{Sec:Conclusions}.

\section{Potentials and Field Excursions} \label{Sec:PotentialsExcursion}
The C1 conjecture asserts a minimum value for the scaled slope, $\lambda$, or curvature, $c^2$, of the potential.  The limiting cases which provide the least deviation from the successful $\Lambda$CDM cosmology are the potentials that keep either of the two parameters constant across cosmic history.

For C1.1, assuming $\lambda$ is constant and a single field model leads to an exponential potential:
\begin{align} \label{Eq:ExponentialPotential}
V(\phi) = A \exp(-\lambda \phi) \,,
\end{align}
where $A$ is the scale of the potential.
Notice that this potential always fails condition C1.2.  
However because C1 can be satisfied either through C1.1 or C1.2, the exponential potential can still be compatible with C1. 

For C1.2, assuming $c^2$ is constant leads instead to a  cosine potential:
\begin{align} \label{Eq:CosinePotential}
V(\phi) = B \cos( c \phi) \,,
\end{align}
where $B$ is the potential scale, and we have not considered an additional overall phase because it does not influence the cosmological evolution. Notice that this potential always fails condition C1.1 but can still be compatible with C1.

Both classes of potential are well motivated from string theory. Supersymmetric models naturally tend to give potentials of this type. For example, race-track models with superpotentials involving multiple gaugino condensates give rise to both classes of potentials. However, it is not unreasonable to expect the low-energy physics to only involve potentials of type Eq.~(\ref{Eq:CosinePotential}) for several axions with other modes massed up at a high scale. Models with of  $O(100)$ axions with potentials that consist of sums of cosines like Eq.~(\ref{Eq:CosinePotential}) can lead to complicated and rich potential landscapes, which are still amenable to analysis~\cite{Bachlechner:2018gew, Bachlechner:2017hsj}. In this work, we will restrict to the simplest case of a single field model.  

The C1 conjecture in either form excludes the $\Lambda$CDM cosmology since it is recovered only for a flat potential where $\lambda=c=0$.
In addition, the second conjecture, C2, when paired with C1 provides an interesting internal tension with cosmology~\cite{Agrawal:2018own}.  
Given a potential with a finite first derivative, the field must roll by at least a finite amount during the past cosmological history.  A large second derivative would also generally imply a finite first derivative except for certain finely tuned initial conditions.   

To calculate the amount of roll during the past expansion history, consider the cosmological Klein-Gordon equation for the field $\phi(N)$, where  $N\equiv \ln a$ is the $e$-folds of the expansion: 
\begin{align}\label{Eq:KG}
\phi'' + \left( 3 + \frac{H'}{H} \right) \phi' + \frac{1}{H^2}\frac{d V}{d \phi} =0 \,.
\end{align}
The primes represent derivatives with respect to the argument, $N$, and $H\equiv  dN/dt$ is the  Hubble parameter which damps the evolution of the field.

Assuming that $\phi'$ is finite at $N\rightarrow -\infty$, Eq.~(\ref{Eq:KG}) has the implicit solution
\begin{equation} \label{Eq:FieldDerivative}
\phi'(N) =  -\frac{e^{-3N}}{H} \int_{-\infty}^N d\tilde N \frac{ e^{3\tilde N}}{H}\frac{d V}{d \phi} \,,
\end{equation}
so that the total field excursion can be written as:
\begin{equation} \label{Eq:TotalFieldExcursion}
\Delta\phi = - \int_{-\infty}^0 dN  \frac{e^{-3N}}{H} \int_{-\infty}^N d\tilde N \frac{ e^{3\tilde N}}{H}\frac{d V}{d \phi} \,.
\end{equation}

In general the total field excursion depends on the potential.  The minimum amount of excursion comes from potentials where the field is nearly frozen by Hubble drag in the radiation and matter dominated epochs and only released during the final e-folds of the expansion during the acceleration epoch.  
These models are known as thawing models. In this case, given the tight current observational constraints in the acceleration epoch, it is usually a good approximation to assume that $V'(\phi)= {\rm const.}$ and  that $H(N)$ can be approximated by the flat $\Lambda$CDM expansion history.
We can then integrate Eq.~(\ref{Eq:FieldDerivative}) and rewrite this in terms of $\lambda$ evaluated around the thawing epoch:
\begin{align} \label{Eq:AppFieldExcursion}
|\Delta \phi| = \frac{\lambda}{3}  \left[\frac{1}{\sqrt{\Omega_\Lambda}}\ln \left( \frac {1+\sqrt{\Omega_\Lambda}}{1-\sqrt{\Omega_\Lambda}} \right)-2 \right] \,.
\end{align}
Here $\Omega_\Lambda=\rho_{\Lambda}/\rho_{\rm tot}$ is the fraction of the total energy density today in $\Lambda$ for 
the assumed $\Lambda$CDM expansion history. 
While this approximation represents a linearization in a small  $\lambda$ around $\Lambda$CDM such that for the scalar field DE $\lim_{\lambda\rightarrow 0}\Omega_{\rm DE}=\Omega_\Lambda$, we shall see that this approximation works across the whole range allowed by the data for an exponential potential. 
This is because of a cancellation between the nonlinearity of the roll and the $\Omega_{\rm DE}(\lambda)$  required by CMB data.   
Therefore, when applying Eq.~(\ref{Eq:AppFieldExcursion}) below, we shall always employ $\Omega_\Lambda=0.69$, which is the best fit for $\Lambda$CDM.
This results in the linear relation 
\begin{equation}
    |\Delta\phi|\approx 0.29 \lambda \,,
    \label{Eq:ApproxFieldExcursion}
\end{equation}
which is  steeper than the one reported in~\cite{Agrawal:2018own} of $|\Delta\phi|\approx \lambda \Omega_{\rm DE}/3 \approx 0.23 \lambda$ by a small, but as we shall see below, significant amount.

Note that the same formula allows us to compute the roll between any two epochs as well.    
To compute $|\Delta\phi|$ from $-\infty$ to some other epoch $N$, we simply make the replacement
\begin{equation}
\Omega_\Lambda \rightarrow \Omega_\Lambda(N)=
\frac{\Omega_\Lambda(0)}{\Omega_\Lambda(0) +  [1-\Omega_\Lambda(0)]e^{-3N}} \,,
\end{equation}
which is the fraction of the total density in the cosmological constant.  We can then take differences of these computations to find the roll between any two epochs that are well after radiation domination.
 
Since thawing models produce the least amount of field excursion, they provide the most incisive combination of
the C1 and C2 conjectures since the field must roll by at least some minimal amount for a given $\lambda$ for 
C1.1 and $\lambda(c,\phi(N\sim 0))$ for C1.2.  
However if we consider C2 alone, then we require a more general relationship between the field excursion and cosmological observables. 
For any canonical scalar field dark energy, we can express
\begin{align} \label{Eq:FieldExcursionExpansionHistory}
|\phi' | = \sqrt{(1+w_{\rm DE}) \frac{\rho_{\rm DE} }{H^2}} \,,
\end{align}
where $w_{\rm DE}=p_{\rm DE}/\rho_{\rm DE}$ is the equation of state parameter for the dark energy.
Assuming that the rest of the matter is in CDM and the known standard model particles, we can
in principle infer $w_{\rm DE}$ and $\rho_{\rm DE}$ from expansion history measurements that
determine $H(N)$ and then integrate $\phi'$ to find the field excursion within the well-measured e-folds.

Current observations are not yet sufficiently precise to fully reconstruct $H(N)$ during
the acceleration epoch without prior assumptions on its functional form, or equivalently the
functional form of $V(\phi)$.  We can however use  reconstruction techniques with very weak priors, 
as in~\cite{Zhao:2017cud}, to constrain the DE equation of state as a function of time. 
This can then be converted to scalar field quantities and in particular the field excursion using Eq.~(\ref{Eq:FieldExcursionExpansionHistory}) (see e.g.~\cite{Huterer:1998qv}) once we impose that $w_{\rm DE}\geq -1$.
We will also use this reconstruction to study the robustness of our conclusions on C1 from the two limiting cases to a generic form of the potential.

In the reconstruction approach, field excursions can be directly computed only between epochs where we have precision distance measurements.
In this context,  we consider only field excursion between redshifts $z=0$ and $z=1.5$ since the latter roughly coincides with the maximum redshift of available supernovae measurements.

\section{Method and data sets} \label{Sec:MethodData}

To test the swampland conjectures discussed in the previous section we will use several cosmological data sets.

As a baseline we use the measurements of the CMB temperature and polarization power spectra at small angular scales from the {\it Planck} satellite~\cite{Ade:2015xua, Aghanim:2015xee} with the addition of the large scale $TEB$ measurements.
We add the {\it Planck} 2015 full-sky lensing potential power spectrum reconstruction~\cite{Ade:2015zua} in the multipole range $40\leq \ell \leq 400$. 
We denote the data set combining these three probes as CMB.

To highlight the power of distance-redshift measurements in testing these conjectures we consider the 
Pantheon Supernovae sample~\cite{Scolnic:2017caz}, that we denote as the SN data set, and 
distance-ladder measurement of the Hubble constant from~\cite{Riess:2018byc}, that we indicate as the $H_0$ data set.

When combining all cosmological data sets together, for completeness, we also employ the following data:
the CMB temperature spectrum measurements at small angular scales from the South Pole Telescope~\cite{2014ApJ...782...74H}
for multipoles  $\ell \le 2500$;
the measurements of the galaxy weak lensing shear correlation function as provided by the Canada-France-Hawaii Telescope Lensing Survey (CFHTLenS)~\cite{Heymans:2013fya} with ultra-conservative cuts~\cite{Joudaki:2016mvz} that make CFHTLenS data insensitive to the modeling of non-linear scales;
measurements of the Baryon Acoustic Oscillation (BAO) scale from BOSS DR12~\cite{Alam:2016hwk}, the SDSS Main Galaxy Sample~\cite{Ross:2014qpa} and 6dFGS~\cite{Beutler:2011hx}.

To produce cosmological predictions and compare them to data, we use the EFTCAMB and EFTCosmoMC codes~\cite{Hu:2013twa,Raveri:2014cka,Hu:2014oga}, modifications to the Einstein-Boltzmann code CAMB~\cite{Lewis:1999bs} and the Markov Chain Monte Carlo (MCMC) code CosmoMC~\cite{Lewis:2002ah} respectively, implementing the quintessence models involved in testing the swampland conjectures.
The quintessence module will be made publicly available in the next release of the EFTCAMB code.

For the  parameters of the quintessence models, we take priors that are flat in the given parameter across a range that is as uninformative as possible.   
In each case we include the standard 6 parameters of the
$\Lambda$CDM model: baryon density $\Omega_b h^2$,   cold dark matter density $\Omega_c h^2$, scalar amplitude $A_s$ and tilt $n_s$, optical depth to reionization $\tau$ and the angular size of the sound horizon $\theta_s$.
We also include all the recommended parameters and priors describing systematic effects in the data sets.
We fix the sum of neutrino masses to the minimal value~(e.g.~\cite{Long:2017dru}).

For the exponential potential, we supplement these parameters with an additional one, $\lambda$, which is allowed to vary in the range $[0,10]$.  
Note that the potential scale $A$ and the initial field position are degenerate and both are absorbed into $\theta_s$. For both the exponential and cosine models, Hubble friction at early times is so large for the allowed cosmological parameters that arbitrary initial kinetic energy is rapidly dissipated and the field effectively reaches the frozen  state after a small number of e-folds. For this reason the initial kinetic energy of the field is not a relevant parameter for either models.

For the cosine potential, we have two parameters: $c$ and the amplitude of the cosine.  We vary $c$ in the range $[0.001,10]$. The lower bound on $c$ is taken to be much less than values of interest for C1.2.   Its presence improves convergence of the Monte Carlo sampling of the posterior, but on its own does not  affect our conclusions.   For $c\ll 1$, the potential is so close to flat that the amplitude is unconstrained.   The upper bound is chosen to be uninformative when paired with the restrictions on the amplitude which we now discuss.

We rescale the cosine amplitude in units of DE density $V_0 =  B/\rho_{\rm DE}(0)$ so that $V_0$ is allowed to vary in the range $[1.001,10]$.
The upper bound is taken to be sufficiently high that it is 
uninformative for $c\ge 0.001$ once all data are considered.
Since energy is covariantly conserved, values of $V_0<1$ are not possible if the field begins at rest.  We impose a slightly higher limit  to remove a special fine tuned case that would avoid observational constraints, at least at the background level.
If the initial field value is set at or very close to the peak, then it will remain there and be indistinguishable from a cosmological constant.

\begin{table*}[!ht]
\setlength{\tabcolsep}{11pt}
\centering
\begin{tabular}{@{}llllll@{}}
\toprule
\multirow{2}{*}{Data set} &  \multicolumn{1}{c}{$\lambda$} & \multicolumn{1}{c}{$|\Delta\phi|\,\,\,[M_P]$} & \multicolumn{1}{c}{$|\Delta\phi|_{z=1.5}\,\,\,[M_P]$} & \multirow{2}{*}{$P(\lambda>1)$} \\
& 68\% (95\%) C.L. & 68\% (95\%) C.L. & 68\% (95\%) C.L. & \\
\toprule
CMB & $\lambda < 1.1\,(1.9)$ & $|\Delta\phi| < 0.33\,(0.52)$ & $|\Delta\phi|_{z=1.5} < 0.29\,(0.45)$ & $38$ \% ($0.9\, \sigma$)\\
\colrule
CMB + SN & $\lambda < 0.38\,(0.64)$ & $|\Delta\phi| < 0.11\,(0.19)$ & $|\Delta\phi|_{z=1.5} < 0.10\,(0.17)$ & $0.017$ \% ($3.8\, \sigma$)\\
\colrule
CMB + $H_0$ & $\lambda < 0.29\,(0.56)$ & $|\Delta\phi| < 0.08\,(0.16)$ & $|\Delta\phi|_{z=1.5} < 0.07\,(0.15)$ & $0.008$ \% ($3.9\, \sigma$)\\
\colrule
ALL & $\lambda < 0.28\,(0.51)$ & $|\Delta\phi| < 0.08\,(0.15)$ & $|\Delta\phi|_{z=1.5} < 0.07\,(0.14)$ & $<0.0006$ \% ($4.5\, \sigma$)\\
\botrule
\end{tabular}
\caption{ \label{Table:ResultsExponential}
The marginalized constraints on the parameters of the exponential potential relevant for the C1.1 and C2  swampland conjectures, for different cosmological data set combinations.
}
\end{table*}

This is an unstable equilibrium and at some point even cosmological perturbations will destabilize it.
To avoid such an unphysical situation we take $V_0\ge 1.001$, which corresponds to forbidding an initial phase $c \phi_i < 0.0447$.
For a random phase $[0,\pi)$, which accounts for reflection symmetry about the origin, this corresponds to about $1.4\%$ of the parameter space but note that our prior
is flat in the range defined by $V_0$. 
We verify that variations around these two cuts do not impact the final results presented later so long as the
priors are taken to be flat in $c,V_0$.

For the choice of the weak prior for reconstruction we follow the quintessence discussion in~\cite{Raveri:2017qvt}.  
We highlight here that, in the redshift range that we use to report field excursion results $z\in [0,1.5]$, the equation of state of DE is allowed by the prior to freely oscillate four times around its mean while faster variations are disfavored by the prior.

\section{Results} \label{Sec:Results}
\begin{figure}[!h]
\centering
\includegraphics[width=\columnwidth]{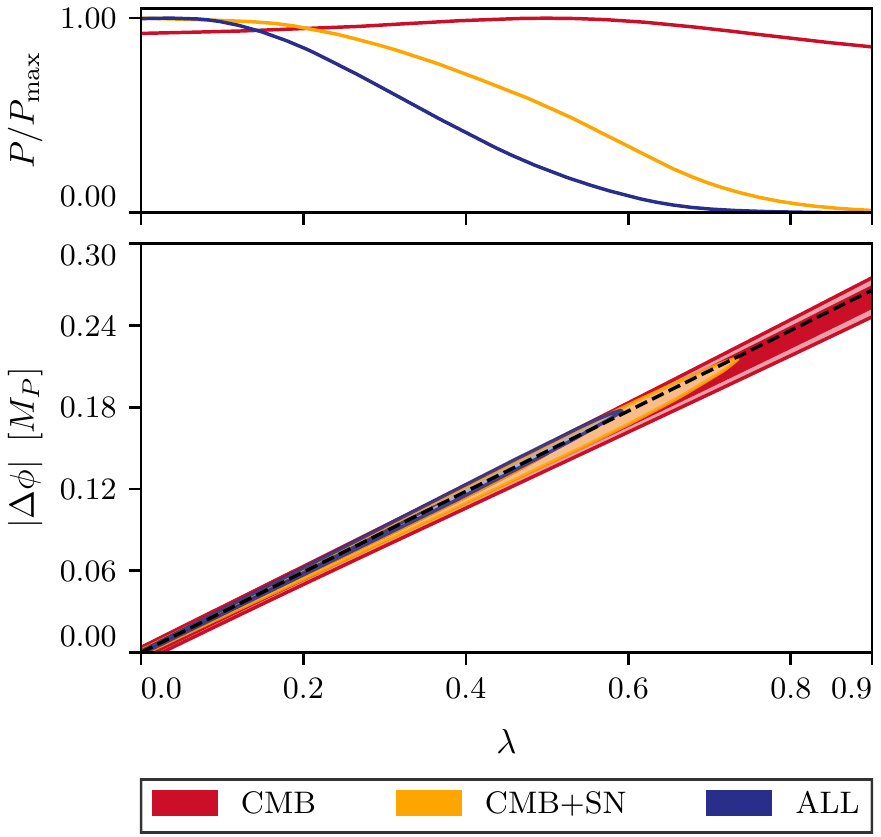}	
\caption{
The marginalized probability distribution of the parameter $\lambda$ of the exponential potential relevant for the C1.1 dS conjecture and the 
joint marginalized distribution of $\lambda$ and total field excursion relevant for the C2 distance conjecture.
The dashed line is the relation between these two parameters predicted by Eq.~(\ref{Eq:ApproxFieldExcursion}).
The darker and lighter shades correspond respectively to the 68\% C.L. and the 95\% C.L. regions.
}
\label{Fig:ExponentialPosterior}
\end{figure}
We first discuss the results for the exponential potential and their implications for C1.1.
The cosmology of the exponential potential is characterized by the field starting deep in radiation domination, frozen in a position in field space by Hubble drag. As Hubble friction decreases at late times the field ``thaws" and starts to roll down the potential, gaining kinetic energy and raising the dark energy equation of state $w_{\rm DE}$, in tension with data in the acceleration regime. 
The $\Lambda$CDM model is recovered only as $\lambda\rightarrow0$, which is inconsistent with C1.1.

As we can see from Table~\ref{Table:ResultsExponential}, when testing the exponential model with CMB only observations, the constraints on $\lambda$ allow ${\cal O}(1)$ values, compatible with C1.1,  as a result of the geometric degeneracy between $\Omega_{\rm DE}$ and $\lambda$ at a fixed distance to recombination required by the measurements.
It is possible to offset distance changes due to a large value of $\lambda$ by lowering the value of $\Omega_{\rm DE}$ which then lowers the Hubble constant.
For this reason, when we combine CMB measurements with direct measurements of the Hubble constant, which prefer a value that is even larger than the one required for $\Lambda$CDM, we strongly constrain the parameter $\lambda$ as a result of the tension between the two measurements.

This effect is also driving some of the constraints in the literature, and should be born in mind when 
interpreting results, especially should the $H_0$ tension be resolved by currently unknown systematics.
Our analysis differs from ~\cite{Agrawal:2018own,Heisenberg:2018yae,Akrami:2018ylq} because we consider all available data sets and examine the robustness of results to various combinations.  These include tests both at the level of the cosmological background and at the level of perturbations; we do not include any additional approximations in the cosmological treatment, nor in extracting model constraints from proxy parameterizations for $w_{\rm DE}(N)$.
Our results on C1.1 are in general qualitative agreement with the results of~\cite{Agrawal:2018own,Akrami:2018ylq}.

Note that the tension with $H_0$ measurements is generic to thawing models, or more generally those quintessence models where the physics at recombination is unmodified from $\Lambda$CDM.  The CMB then constrains the physical matter density $\rho_m$ and distance to $z_*$, the redshift of  recombination, $D_*=\int_0^{z_*} dz/H$ directly.
Since $w_{\rm DE}\ge -1$, the dark energy can only redshift faster than a cosmological constant. Therefore, for a fixed distance, its contribution to the present energy density must be smaller, and hence $H_0^2 \propto \rho_m(0)+\rho_{\rm DE}(0)$ must also be smaller. 

\begin{table*}[!ht]
\setlength{\tabcolsep}{11pt}
\centering
\begin{tabular}{@{}llllll@{}}
\toprule
\multirow{2}{*}{Data set} &  \multicolumn{1}{c}{$c$} &
\multicolumn{1}{c}{$\lambda_{\rm eff}$} &
\multicolumn{1}{c}{$|\Delta\phi|\,\,\,[M_P]$} & \multicolumn{1}{c}{$|\Delta\phi|_{z=1.5}\,\,\,[M_P]$} & \multirow{2}{*}{$P(c>1)$} \\
& 68\% (95\%) C.L. & 68\% (95\%) C.L. & 68\% (95\%) C.L. & 68\% (95\%) C.L. & \\
\toprule
CMB & $c < 2.3\,(3.1)$ & $\lambda_{\rm eff} < 1.4\,(2.2)$ & $|\Delta\phi| < 0.51\,(0.66)$ & $|\Delta\phi|_{z=1.5} < 0.47\,(0.63)$ & $50$ \% ($0.6\, \sigma$)\\
\colrule
CMB + SN & $c < 0.25\,(1.4)$ & $\lambda_{\rm eff} < 0.40\,(0.71)$ & $|\Delta\phi| < 0.11\,(0.19)$ & $|\Delta\phi|_{z=1.5} < 0.10\,(0.16)$ & $8.5$ \% ($1.7\, \sigma$)\\
\colrule
CMB + $H_0$ & $c < 0.17\,(0.84)$ & $\lambda_{\rm eff} < 0.31\,(0.58)$ & $|\Delta\phi| < 0.09\,(0.16)$ & $|\Delta\phi|_{z=1.5} < 0.08\,(0.15)$ & $3.3$ \% ($2.1\, \sigma$)\\
\colrule
ALL & $c < 0.16\,(0.73)$ & $\lambda_{\rm eff} < 0.29\,(0.53)$ & $|\Delta\phi| < 0.08\,(0.15)$ & $|\Delta\phi|_{z=1.5} < 0.07\,(0.14)$ & $1.9$ \% ($2.3\, \sigma$)\\
\botrule
\end{tabular}
\caption{ \label{Table:ResultsCosine}
The marginalized constraints on the parameters of the cosine potential relevant for the C1.2 and C.2 swampland conjectures, for different cosmological data set combinations.
}
\end{table*}

Even though CMB+$H_0$ data provide the largest component of the overall constraint,  large values of $\lambda$ are also disfavored by CMB and supernovae measurements.
Since the SN likelihood is marginalized over an overall calibration, it does not constrain the Hubble constant but rather the shape of the distance redshift relation $D(z)$.  
This makes  the conclusion that $\lambda \sim O(1)$ is disfavored by cosmological observations more robust, as it comes from both the normalization and shape of $D(z)$.

As we combine all data sets together the results only tighten slightly compared with the CMB+$H_0$ constraint.
The probability of exceeding the value of $\lambda=1$ parallels this trend, as can be seen from Table~\ref{Table:ResultsExponential}, and becomes negligible as we combine CMB observations with low redshift distance measurements, reaching a value equivalent to a $4.5\sigma$ exclusion with the ALL data set combination.  
We compute here the effective number of standard deviations that we would associate to an event of given probability as
$n_{\sigma} \equiv \sqrt{2}{\rm Erf}^{-1}(1-P)$ to aid the interpretation of the statistical significance of the reported results.  Thus our estimates in highly excluded regions is limited by the finite MCMC sample.

Since the exponential case corresponds to a thawing model, the total field excursion converges over the whole cosmological evolution and we report its upper bound in Table~\ref{Table:ResultsExponential}.   
From Figure~\ref{Fig:ExponentialPosterior}, we see that it is  tightly correlated with $\lambda$ as predicted by Eq.~(\ref{Eq:ApproxFieldExcursion}).
Note that if we use the slope reported in~\cite{Agrawal:2018own}, the small
difference is highly significant due to the tight correlation between the two parameters imposed by the data. 
Interestingly, our linear prediction is also robust to $\lambda \sim {\cal O}(1)$, where we would expect to have non-linear corrections to Eq.~(\ref{Eq:AppFieldExcursion}) because they are partially compensated by the change in $\Omega_{\rm DE}(\lambda)$ required to fix the distance to recombination.
As we can clearly notice in Figure~\ref{Fig:ExponentialPosterior}, the correlation between these two parameters is set by CMB observations which define the geometric degeneracy direction.  The allowed width orthogonal
to this direction reflects the small uncertainty on the distance to recombination, while the extent of 
the degeneracy is limited by the data in the acceleration regime.

Given that C1.1 with $\lambda \gtrsim 1$ is ruled out by observations, we now turn to whether the dS conjecture can instead be satisfied through C1.2 using the cosine potential.

The cosine model also falls into the class of thawing models.
To provide the necessary ingredient to drive cosmic acceleration the field  has to start its evolution deep in radiation domination close to the positive maximum of the potential, where it is frozen by Hubble friction.
As Hubble drag relaxes the field starts rolling down across a region in potential that corresponds to a tachyonic mass.
A tachyonic $|m| \gtrsim H_0$ would generally cause this rolling to violate observational constraints on acceleration.

Similarly to the exponential potential, we can see from Table~\ref{Table:ResultsCosine}, that CMB only observations would allow very large values of $c$ as a result of the geometric degeneracy.   
On the other hand combining CMB measurements with distance-redshift data disfavors large values of $c$ since they generally imply a substantial deviation from $w_{\rm DE}=-1$.

The probability that $c$ exceeds one follows the same qualitative behavior as $\lambda$ in the exponential potential and falls as we add the distance-redshift data.
Notice that the distribution of $c$ is highly non-Gaussian because of a degeneracy between $V_0$ and $c$ in
determining the dark energy equation of state or equivalently, as we shall see, the local slope of the potential. This implies that the 95\% C.L. bound is significantly larger than twice the 68\% C.L. one.
\begin{figure}[!ht]
\centering
\includegraphics[width=\columnwidth]{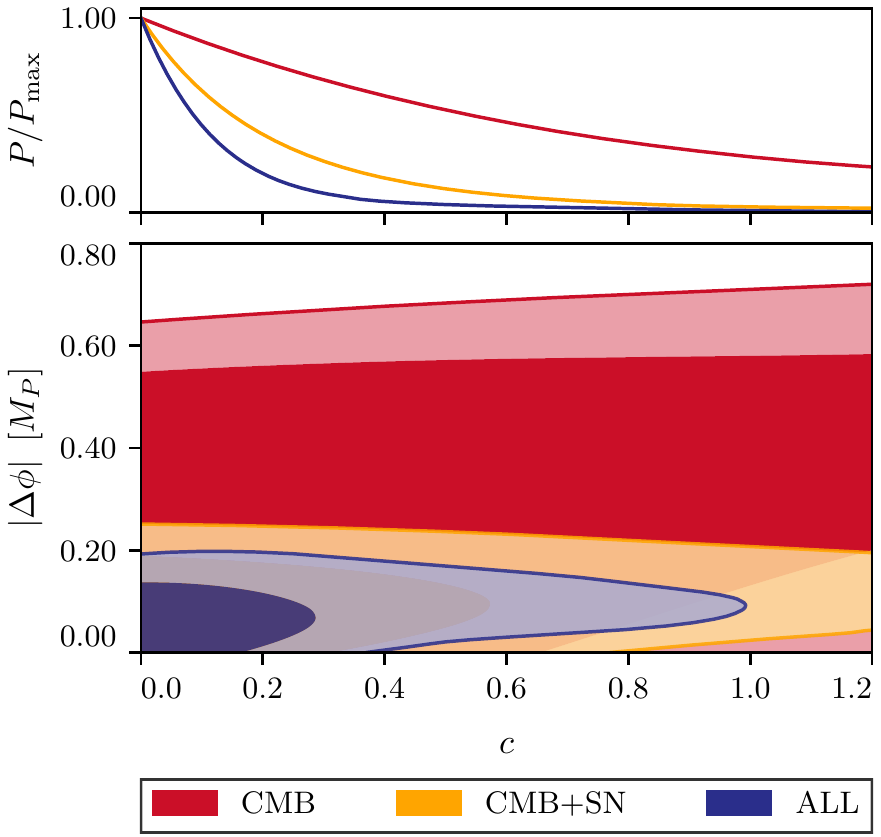}	
\caption{
The marginalized probability distribution of the parameter $c$ of the cosine potential, relevant for the C1.2 dS conjecture, and the joint marginalized distribution of $c$ and total field excursion relevant for the C2 distance conjecture.
The darker and lighter shades correspond respectively to the 68\% C.L. and the 95\% C.L. regions.
}
\label{Fig:CosinePosterior}
\end{figure}

This can also be clearly appreciated in the upper panel of Figure~\ref{Fig:CosinePosterior}.
In the lower panel we show the joint marginalized posterior of the parameter $c$ and total field excursion which are almost uncorrelated once data in the acceleration epoch is included.

In spite of this lack of correlation, the cosine model still falls into the thawing class where Eq.~(\ref{Eq:AppFieldExcursion}) holds.   
The lack of correlation reflects the ability for a single value of $c$ to take on different values for the local slope of the
potential. 
We extract the slope of the potential at the thawing epoch by averaging $\lambda(N)$ for the cosine potential and weighting it by $\Omega_\Lambda(N)$ from the best fit $\Lambda$CDM model. 
We verify that other choices do not result in appreciable differences.
We refer to the resulting quantity as $\lambda_{\rm eff}$ that should serve as a proxy for the $\lambda$ in Eq.~(\ref{Eq:AppFieldExcursion}).

\begin{figure}[!ht]
\centering
\includegraphics[width=\columnwidth]{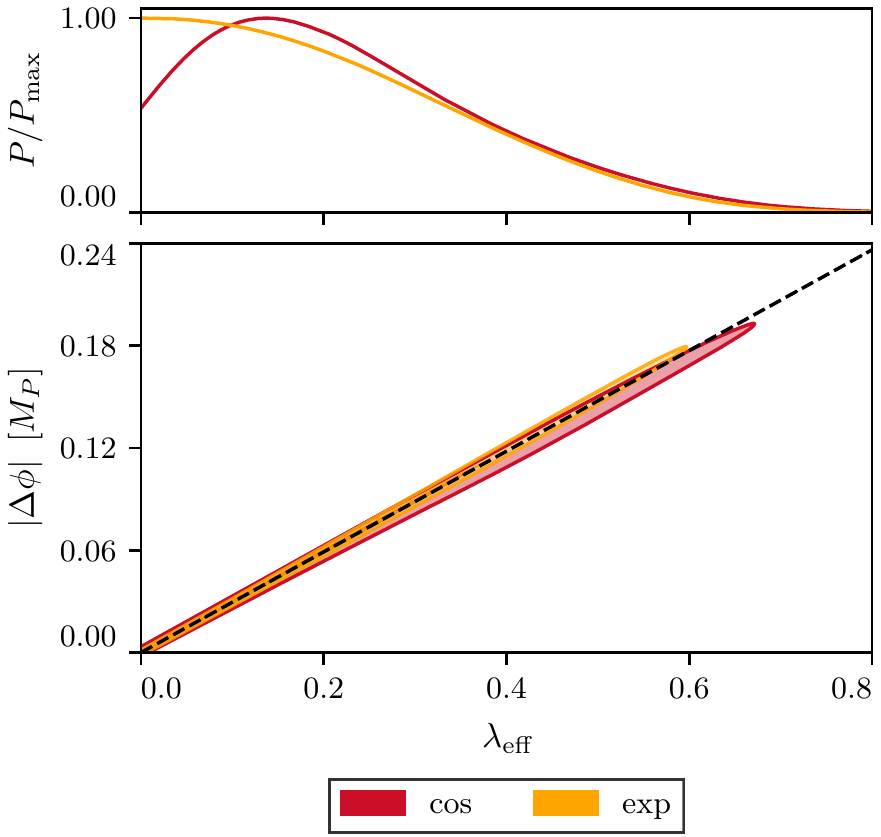}	
\caption{
The marginalized probability distribution of $\lambda_{\rm eff}$ for both the exponential and cosine potentials together with its joint marginalized distribution with total field excursion for the ALL dataset.
The dashed line is the relation between these two parameters predicted by Eq.~(\ref{Eq:ApproxFieldExcursion}). 
The darker and lighter shades correspond respectively to the 68\% C.L.\ and the 95\% C.L.\ regions.
}
\label{Fig:LambdaEffPosterior}
\end{figure}

In Figure~\ref{Fig:LambdaEffPosterior} we show the joint marginalized posterior of $\lambda_{\rm eff}$ and total field excursion. These two parameters are now strongly correlated and follow almost exactly the relation in Eq.~(\ref{Eq:ApproxFieldExcursion}) written in terms of $\lambda_{\rm eff}$.
The limits imposed by the data also agree well between the exponential and the cosine models as we can see by comparing Table~\ref{Table:ResultsCosine} and Table~\ref{Table:ResultsExponential}.
The difference near $\lambda_{\rm eff}=0$ reflects the fact that initial conditions where the field starts at the top of the cosine potential require fine tuning, and are downweighted with our choice of priors.

Upper limits on $\lambda_{\rm eff}$ are robust  because in thawing models, observations mainly constrain one parameter: $\lambda$ at the thawing epoch, for which $\lambda_{\rm eff}$ is a proxy. 
This also explains why the marginal distribution of $c$ in Fig.~\ref{Fig:CosinePosterior} is so non-Gaussian and leads to weaker constraints on the dS conjecture C1.2 for the cosine than constraints
on C1.1 for the exponential potential.  The physical reason for this is as follows: sufficiently close to the peak of the cosine potential, it becomes indistinguishable from a cosmological constant at the background level, even for large values of $c$.

To estimate the amount of tuning required to allow a given value of $c$ we can use the constraints on $\lambda$ from the exponential potential.
In the cosine model:
\begin{align} \label{Eq:tuning}
\lambda_{\rm eff} =&\, c \tan ( c \phi) \approx  c \tan (c(\phi_i+\Delta\phi)) \nonumber\\
\approx &\, c \tan (c (\phi_i + 0.29\lambda_{\rm eff})) \,,
\end{align}
where $\phi_i$ is the initial field position and we have
employed Eq.~(\ref{Eq:ApproxFieldExcursion}) to estimate the amount of roll from the intial value.
We can now take constraints for $\lambda$ from the exponential potential, leverage on the fact that for the cosine potential constraints on $\lambda$ and $\lambda_{\rm eff}$ are very close, and invert Eq.~(\ref{Eq:tuning}) to obtain the amount of initial condition tuning needed for a given $c$:
\begin{align} \label{Eq:TuningExplicit}
c \phi_i = -0.29\, c\, \lambda_{\rm eff} +\arctan\left( \frac{\lambda_{\rm eff}}{c} \right) \,.
\end{align}
As an example, if we take the 95\% C.L. bound from the ALL dataset $\lambda_{\rm eff}\approx\lambda <0.51$, we would require $c\phi_i/\pi <0.1$ for $c>1$.

\begin{figure}[!ht]
\centering
\includegraphics[width=\columnwidth]{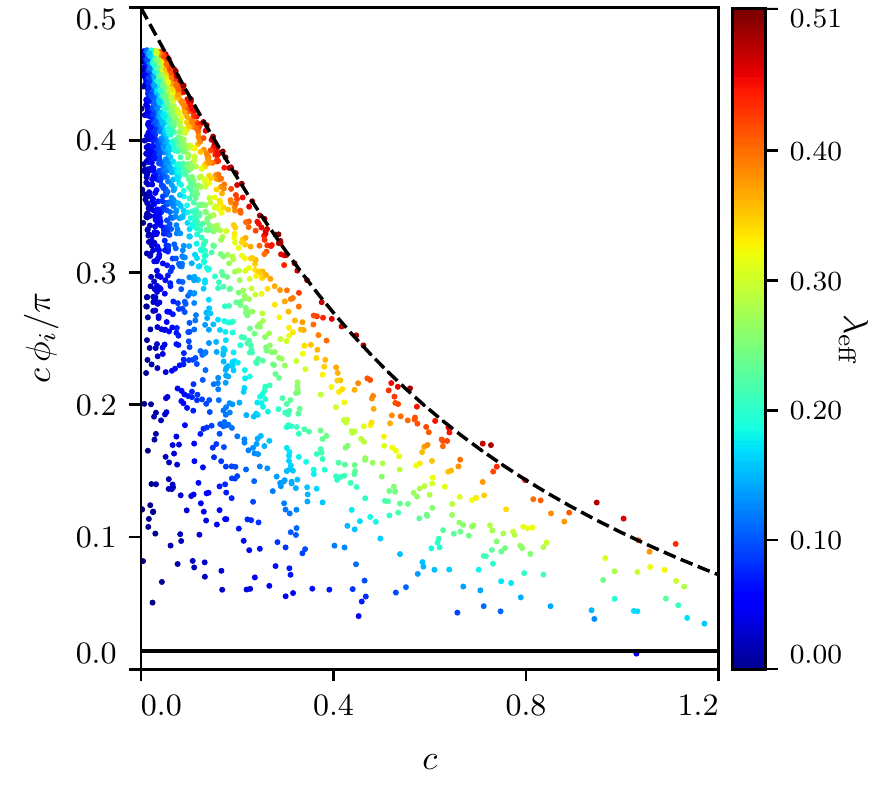}	
\caption{
The joint marginalized distribution of initial condition tuning and $c$ for the cosine potential and the ALL dataset.
Models are cut based on their value of $\lambda_{\rm eff}$ at the $95\%$ C.L. bound from the exponential potential, resulting in models shown with $\lambda_{\rm eff}<0.51$.
The density of points is proportional to the joint PDF and the color represents the value of $\lambda_{\rm eff}$.
The dashed line represents the amount of tuning needed to stabilize a given value of $c$ given by Eq.~(\ref{Eq:TuningExplicit}).
The solid line represents the tuning cut that we enforce.
}
\label{Fig:Tuning}
\end{figure}

This effect is clearly seen in Figure~\ref{Fig:Tuning} where we show the joint distribution of initial condition tuning and $c$ as a cloud of points, colored by their value of $\lambda_{\rm eff}$ and cut at the 95\% C.L. bound on $\lambda$ of the ALL data set for the exponential potential.
We can see that Eq.~(\ref{Eq:TuningExplicit}) matches  the 95\% C.L. bound very well.
This further justifies the use of the exponential model constraint on $\lambda$ to estimate the tuning of the cosine model in Eq.~(\ref{Eq:TuningExplicit}).
For large values of $c$ there is a small discrepancy that is due to non-linearities in $\lambda_{\rm eff}$ in the evolution of field roll.

Extreme values of $c$ are then downweighted by two effects:
our flat prior on $V_0$ gradually disfavours tuned solutions and a hard tuning cut at $c \phi_i / \pi = 0.014$ avoids extreme values that would slow the convergence of the parameter estimation chains.

Although our results are robust to the prior ranges for
the ALL data set, they do depend on the shape of the
prior.  Were there to be a physical reason to favor the tuned cases where the field remains stuck at the top of the potential, then larger values of $c$ would be allowed by the data,
as quantified in Figure~\ref{Fig:Tuning}.  Conversely, were there some reason that the prior should be flat in $\log c$, then the posterior bounds on $c$ would tighten.
For any given choice of prior,
our technique of adopting the exponential potential constraint on $\lambda_{\rm eff}$ provides a simple means
of estimating implications for $c$.

We conclude that C1.2 with $c>1$ is also disfavored by the data, except for fine tuned and unstable initial conditions.   
Hence the data is in tension with both versions of the C1 dS conjecture. 

Next we investigate the robustness of these results to allowed changes in the potential obtained by reconstructing $w_{\rm DE}(N)$ from the data.
We first remark that even allowing an arbitrary potential, there is no significantly better fit to the data than the $\Lambda$CDM model.

In the reconstruction, where  both $\lambda$ and $c$ become time dependent, we extract their minimum value to assess C1 and compare their constraint to the limiting cases.
For both $\lambda$ and $c$, we find that the constraints from reconstruction are tighter for both parameters, making the two limiting potentials the most conservative assessment of the dS conjecture.
In both cases the reason is as follows: at a given $\lambda_{\rm min}$, the field will generally cross into regions of larger $\lambda$ that would result in larger deviations from $\Lambda$CDM.
As derived from the general reconstruction these extra deviations are not favored by the data making the model with a given $\lambda_{\rm min}$ more, or at least equally, disfavored with respect to the exponential with $\lambda = \lambda_{\rm min}$.
A similar argument can be made for $c$ and the cosine potential.

Finally we consider field excursion from reconstruction.  
Unlike for the thawing class of models, reconstruction allows potentials where the field rolls significantly at high redshift. This is not as well constrained by the data and so we
focus on the amount of roll between $z=1.5$ and the present.
The corresponding constraints for the exponential and cosine potentials are reported in Table~\ref{Table:ResultsExponential} and Table~\ref{Table:ResultsCosine} respectively, and follow accurately the behavior given by Eq.~(\ref{Eq:AppFieldExcursion}).  
In these thawing models, the data sets a robust upper bound on the amount of roll between these two epochs,
$\Delta\phi_{z=1.5}< 0.07\,(0.14)$ at 68\%\,(95\%) C.L. that is only slightly smaller than the total excursion. 

For the more general case of reconstruction, we have a weaker upper limit: $\Delta\phi_{z=1.5}< 0.18\, (0.22)$ at 68\%\,(95\%)  C.L.
In the reconstruction no potential is assumed a priori but a smoothness criterion for the equation of state has to be assumed, as in~\cite{Zhao:2017cud} and~\cite{Raveri:2017qvt}.
To understand whether the prior is limiting this determination we run a prior only chain that results in much larger allowed field excursions of $\Delta\phi_{z=1.5}< 1.22\,(1.6)$ at $68\%\,(95\%)$ C.L.
We also verify that the data likelihood decreases as expected between $\Delta\phi_{z=1.5}=0.18$ and $\Delta\phi_{z=1.5}=0.22$ showing that the constraint reflects the preference of the data not the prior. 
As a further check, note that for a constant $w_{\rm DE}$ a bound on $\Delta\phi_{z=1.5}<0.22$ corresponds
to $w_{\rm DE}<-0.95$ which is roughly the level at which such deviations are allowed with current data.

\section{Conclusions} \label{Sec:Conclusions}

In this paper we studied the cosmological implications of the refined de Sitter (C1) and distance (C2) swampland conjectures that have been proposed in literature. The C1 conjecture depends on two dimensionless constants $(\lambda, c)$. 

We have determined which piece of experimental evidence contributes most to data constraints on these conjectures. We found that the strongest constraints are driven by the synergy between CMB observations fixing the distance to recombination, and both the normalization and shape of the distance redshift relation.
The normalization, or Hubble constant, is especially powerful in establishing constraints since these quintessence models exacerbate the already existing tension in $\Lambda$CDM.  

Overall we found that, combining most of the available cosmological data sets,
$\lambda<0.51$ and $c<0.73$ at the 95\% confidence level.
Both results are obtained by directly computing cosmological predictions for the quintessence models involved, without approximations, and properly comparing them to the data.
In this respect the result on $\lambda$ settles the discussion in the literature on the assessment of
C1.1, and extends these results to the complete current refined de Sitter conjecture C1.  Only specially fine-tuned initial conditions, where the field starts at the unstable maximum of the potential, can evade the bound on $c$.

As a benchmark for the tension between these conjectures and cosmological observations, we computed the probability that $\lambda$ and $c$ can exceed one and find that for the most complete data compilation: 
$P(\lambda>1)<0.0006\%$, or equivalently disfavored at a statistical significance higher than $4.5\sigma$ and limited by our sampling of the tails of the distribution; 
$P(c>1)=1.9\%$, or equivalently disfavored at the $2.3\sigma$ level.
Even without the Hubble constant measurements, these results remain significant.

We have also derived a general and accurate relationship between $\lambda$ and field excursion that applies to the whole class of thawing quintessence models.   
For these models, the observations place an upper bound at 95\% C.L. of  $|\Delta \phi|<0.15 M_P$.   

To comment on the robustness of these results to changes in the form of the potential,  we have considered non-parametric reconstructions of the equation of state of DE and its projection on quintessence models.
We have verified that in this general setup, the exponential and cosine potentials are the limiting cases for the two parts of the C1 conjecture.

We discussed the relationship between field excursion and directly observable quantities, and used the reconstruction results to compute field excursion in the observable data range. At 95\% C.L., this results in $|\Delta\phi|_{z=1.5}< 0.22M_P$. 

The field excursion results that we have found exhibit no tension with the distance conjecture, which is the swampland conjecture on the firmest theoretical footing. How one views the results on $(\lambda, c)$ depends on what one considers to be $O(1)$, and the confidence level one is willing to assume. 

At the 95\% C.L., the constraints $(\lambda<0.51, c<0.73)$ are not particularly troubling. One could easily imagine a more precise conjecture emerging from string theory involving dimensionless numbers of that size. At the 68\% C.L. where the constraints take the form $(\lambda<0.28, c<0.16)$, the numbers start to look a little more in need of some theoretical explanation.

\acknowledgments
We thank
Samuel Passaglia and Kimmy Wu
for useful comments.
MR and WH are supported by U.S.~Dept.~of Energy contract DE-FG02-13ER41958.  
WH was further supported by NASA ATP NNX15AK22G and the Simons Foundation.
SS is supported, in part, by NSF Grant No.~PHY-1316960. Computing resources were provided by the University of Chicago Research Computing Center through the Kavli Institute for Cosmological Physics at the University of Chicago. 
\bibliography{biblio}

\end{document}